# High-harmonic generation in amorphous solids

Yong Sing You[1], Yanchun Yin[2,3], Yi Wu[2,3], Andrew Chew[2,3], Xiaoming Ren[2,3], Fengjiang Zhuang[2,3], Shima Gholam-Mirzaei[3], Michael Chini[3], Zenghu Chang[2,3], and Shambhu Ghimire[1]

[1]Stanford PULSE Institute, SLAC National Accelerator Laboratory, Menlo Park, California 94025, USA
[2]Institute for the Frontier of Attosecond Science and Technology, CREOL, University of Central Florida, Orlando, Florida 32816, USA
[3]Department of Physics, University of Central Florida, Orlando, Florida 32816, USA

**Abstract**
High-order harmonic generation (HHG) in isolated atoms and molecules has been widely utilized in extreme ultraviolet (XUV) photonics and attosecond pulse metrology. Recently, HHG has also been observed in solids, which could lead to important applications such as all-optical methods to image valance charge density and reconstruction of electronic band structures, as well as compact XUV light sources. Previous HHG studies are confined on crystalline solids; therefore decoupling the respective roles of long-range periodicity and high density has been challenging. Here, we report the first observation of HHG from amorphous fused silica. We decouple the role of long-range periodicity by comparing with crystal quartz, which contains same atomic constituents but exhibits long-range periodicity. Our results advance current understanding of strong-field processes leading to high harmonic generation in solids with implications in robust and compact coherent XUV light sources.

**Introduction**
High-order harmonic generation (HHG) in isolated atoms[1,2] has been the foundation of attosecond pulse metrology[3], extreme ultraviolet (XUV) photonics[4], and molecular orbital tomography[5]. Now that the strong-field response producing nonperturbative high-order harmonics has also been identified in bulk solids[6–14] there are strong motivations in probing electronic structure of solids including normally unoccupied conduction bands[8,9,15], and to overcome some of the drawbacks of the gas phase XUV sources. Gas phase XUV sources suffer from low efficiency and therefore do not provide sufficient flux desired for many applications such as metrology and imaging[16]. Solid-state HHG has the potential for high efficiency because of the use of rigid and high-density target. Since the original discovery in single-crystal ZnO, several crystalline solids, such as GaSe, $SiO_2$, Ar, Kr, MgO and $MoS_2$, have been used for HHG[6–14]. The important findings such as high-energy cut-off scaling with the field[6], emergence of a secondary plateau[11,17] and complex ellipticity dependence[14], indicate that the underlying electron dynamics are markedly different from the prediction made by the widely accepted 3-step re-collision model for atomic HHG[18]. These fundamental differences are attributed to the high-density and periodicity as the field-driven electron is always in the proximity of Coulomb potential[6,19]. In order to incorporate the fundamental solid-state response two major mechanisms are being considered, which are based on the emission from nonlinear inter-band polarization and intra-band current[19–23]. While this topic is of intense debate at this time both mechanisms rely on the basic assumption of Bloch theorem[24], where the electron is moving in a periodical potential such that its energy eigenstates are Bloch waves. Here, we ask these a fundamental question—can high harmonics be generated from strongly driven amorphous solids, which lack periodicity? The answer would help to elucidate the role of long-range periodicity in high harmonics generation.

We report the first observation of nonperturbative high-order harmonic generation from amorphous silicon dioxide (fused silica) subjected to intense few-cycle laser pulses of field strength ~ 2 VÅ$^{-1}$ without damage. High-harmonic spectrum shows characteristics of both spatial and temporal coherence and the photon energy extends up to ~25 eV. In order to understand the role of long-range periodicity we perform similar measurements in crystalline silicon dioxide (quartz). We compare their generation efficiency, dependency with input field strength, and the laser waveform dependence through the carrier-envelope phase (CEP) setting of the driving laser. While the spectral extent of harmonics is similar, which is up to ~25 eV, the harmonics from crystal quartz shows a minima separating two plateau structures. We reproduce our experimental observation by performing quantum calculations of driven multi-level system, where the energy levels correspond to the respective electronic band structures.

## Results

**High harmonic spectrum from amorphous solids**

In the experiment, we focus two-cycle laser pulses obtained from high-efficiency optical parametric chirped-pulse amplification (OPCPA) system into a 100 μm thick sample of amorphous fused silica to generate high harmonics (see Methods). The laser spectrum is centered ~1700 nm (0.73 eV). The estimated maximum peak field strength inside the samples without damage is ~2 VÅ$^{-1}$. The relatively high damage threshold is achieved with the use of ultrashort pulse duration and long wavelength. For comparative study we introduce crystal quartz sample of similar thickness in the same setup. Figure 1 shows representative high harmonic spectra from fused silica and crystal quartz. In both cases, the spectrum extends to ~25 eV, corresponding to 33$^{rd}$ harmonics. Harmonic spectrum from fused silica consists of odd-order harmonic peaks although the peak position may or may not correspond to the exact harmonic order depending on the CEP setting (see methods). The spectrum from crystal quartz has more peaks because of the inclusion of additional even order harmonic peaks thus forming more continuous pattern (except the dip at ~18 eV). Presence of even harmonics in crystal quartz is consistent to the fact that the crystal structure does not exhibit inversion symmetry[25]. However, fused silica is isotropic and therefore nominally it does not produce even-order harmonics. A unique feature in the spectrum from quartz is the presence of a prominent minimum in the range from 17 to 18 eV, which separates the spectrum into two plateaus. Similar to rare gas solids[11], the origin of multiple plateaus can be attributed to the important role of high-lying conduction bands.

**Scaling of high harmonics with laser peak fields**

We study how the generation from periodic and non-periodic medium depends on the laser field. We scan the laser field strength from 1.4 to 2.0 VÅ$^{-1}$ and the results are shown in Figure 2 **a** and **b**. We find that the spectral minima of crystal quartz does not shift as a function of the laser field. At moderate field strengths harmonic yield from fused silica and crystal quartz are about the same. The harmonic yield increases with the input laser field strength in both cases but the scaling are different such that at the highest peak field crystal quartz is about 4 times more efficient. In quartz, the Si-O bonding directions is periodically aligned with respect to the laser field, in contrast to the random network structure in amorphous fused silica; therefore the expected nonlinearlity is higher in quartz. We note that Luu *et al*[9] have also reported similar enhancement in crystal quartz compared with conventional gas phase atomic targets. Nonetheless it is remarkable that the high-energy cutoff of HHG spectrum from non-periodic samples like fused silica is same as that from perfect crystals.

**CEP dependence of high harmonic spectrum**

To gain insights into underlying electron dynamics we perform time-domain measurements through the CEP dependence[17]. Figure 3 **a** and **b** show the measured CEP scans for fused silica and crystal quartz respectively. It is seen that in fused silica the photon energy of harmonic peaks shift in energy (strips, identified by the dashed line) with a CEP slope ~3eV/π and the harmonic spectrum repeats with π periodicity (horizontal axis). However, the harmonic spectra from crystal quartz show a dominant monotonic 2π periodicity along with some resemblance of strips in the vertical direction. We note that the laser wavelength is much longer than the interatomic distances in solids. Due to the random interatomic structure in fused silica, both positive and negative polarity (half cycle) experience the same average response consistent to the observed π periodicity in the CEP. However, crystal quartz exhibits a broken inversion symmetry and as a result the positive and negative half cycles experience different collective response, resulting in a 2π periodicity. The origin of CEP strips is an indication of atto-chirp, which we discuss below.

**Simulation**

To model our results, we solve time-dependent Schrodinger equations in a multi-level system[22] using density matrix approach (see Methods). For fused silica, we consider a two-level system, where the separation corresponds to the experimental band gap ~9 eV[26]. Figure 3c shows the simulation results, which reproduces the experimentally observed CEP slope/strip and appropriate CEP periodicity. We note that CEP slope is a general feature of few-cycle field driven few-level system. In order to explain the observed minima in harmonic spectrum from crystal quartz we use a simple three-level system similar to previous result[17]. Here, the separation between first and second level corresponds to the primary experimental bandgap (~9 eV) and the separation between second and third level corresponds to the second band gap (~4 eV). Also, in order to account for the non-centrosymmetric structure of crystal quartz we include a permanent dipole moment term (see Methods). The simulation results are shown in figure 3d. We reproduce the dominant monotonic 2π periodicity along with the spectral minima (at around 18 eV) that is independent of the CEP settings. Our CEP measurements are relative, however it can be seen that in our scale the spectrum is maximized when CEP = π, 3π, and 5π, which means that the permanent dipole moment is periodically aligned/anti-aligned to the peak of the laser field.

**Atto-chirp analysis**

As we discuss in the methods section below the CEP slope corresponds to the sub-cycle delay between harmonics, which is known as atto-chirp [see Methods and You et al.[17]]. In the gas phase, the atto-chirp of plateau harmonics were found largely independent of the Coulomb potential[27]. In contrast, we find that the measured CEP slope in fused silica is about 2 to 3 times larger than that in the case of MgO under similar laser parameters[17]. On the other hand crystal quartz shows negligible atto-chirp, where high harmonics are emitted predominantly at specific CEP settings ( π, 3π and 5π). This monotonic CEP dependence (without slope) is consistent with the recent demonstration of chirp-free harmonics through time-domain measurements[12]. In our simulation, such monotonic dependence comes from the non-centrosymmetric structure of quartz, where harmonics are strongly enhanced when laser field is parallel to the permanent dipole moment. Therefore, our analysis indicates that the atto-chirp in solid-state harmonics depends strongly on the translational periodicity in materials.

**Conclusion and Outlook**

We present the discovery of nonperturbative high-order harmonics from non-periodic transparent solids subjected to strong laser fields. The harmonic spectrum exhibits a broad plateau structure, which extends to ~25 eV limited by the damage threshold at ~2 VÅ$^{-1}$. The strong CEP dependence of the HHG spectrum confirms that the harmonics are locked in phase with the driving laser field. The measured CEP slope in fused silica indicates that harmonics are delayed with respect to each other in the sub-cycle scale. This sub-cycle delay could be utilized to probe inter-band tunneling time in strongly-driven solids. The observation of HHG from amorphous materials means that periodicity of atomic arrangement inside solids is not a requirement for generating coherent XUV radiation. When combined with the modest requirements in the peak intensity (~$10^{13}$ W/cm$^2$) the solid-state HHG technique becomes an attractive candidate for future high-repetition rate compact XUV light sources[28]. Amorphous optical materials are readily available and can be incorporated relatively easily in photonics design such as in intra-cavity XUV frequency comb, which currently uses gas targets[29]. Another advantage of using solid targets is that relatively large beam size can be utilized to further enhance the efficiency. Finally, we foresee that HHG in amorphous materials might open up new series of possibilities in nano-photonics and XUV waveguides[30].

**Methods:**

**Experimental setup**

We focus two-cycle laser pulses produced from a high-efficiency optical parametric chirped-pulse amplifier (OPCPA) laser system[31] into 100 μm thick samples placed inside the vacuum chamber. The laser spectrum is centered ~1.7 μm and the pulse duration is ~11 fs, measured by frequency-resolved optical gating. The samples withstand the peak intensity ~ $10^{14}$ W/cm$^2$ (~ 2.1 V/Å) at 1 kHz repetition rate. Such a relatively high damage threshold is reached due to the combination of relatively large band gap (9 eV), small photon energy (0.73 eV) and the ultra-short pulse duration. The CEP settings are adjusted using an acousto-optical modulator, which provides relative values. We record the harmonic spectra with an imaging spectrometer consisting of a flat-field variable groove density grating and microchannel plates. The spectral range is from 13 eV to 25 eV, limited by the collection angle of the XUV spectrometer. The spectrum is not corrected for the sensitivity of the grating.

**Long-range order**

To discuss the role of periodicity for microscopic HHG process, we compare the maximum excursion distance of semi-classical electron with the grain size of sample. The maximum excursion distance of semi-classical electron is $r_{max} = eE\lambda^2/4\pi^2 mc^2$, where E is the electric field and $\lambda$ is the wavelength. At the highest field E= 2 V/Å this corresponds to ~30 Å. To confirm the local randomness in amorphous fused silica we perform x-ray powder diffraction measurements. Based on Scherrer particle size equation[32], the measured width of x-ray diffraction ring (~10 degrees) corresponds to a grain size of about 9 Å, which is an upper bound (resolution limited). Clearly, in fused silica the local correlation lengths are much shorter than the excursion length at E= 2 V/Å. In contrast, crystal quartz is a single-crystal material thus the correlation length is infinitely long compared to the excursion distance. Therefore, at highest peak fields we are comparing amorphous and crystalline medium of same atomic constituents. At moderate fields, such as at E= 0.5 V/Å, the excursion length would be around 8 Å, approaching the measured correlation length in fused silica. Therefore at moderate fields we do not expect significantly different harmonic efficiency between fused silica and crystal quartz, consistent to the experimental results.

**CEP dependence of high harmonics spectrum**

The CEP dependence is modeled by considering the interference between adjacent XUV pulses[17,33]. The XUV bursts generated in each half cycle have different amplitudes and phases that depend on the instantaneous intensity and the CEP, which can lead to constructive or destructive interference[33]. Consider the time-dependent dipole of two adjacent bursts:

$$D(t) = d_1(t) + d_2(t)$$
$$= d(t) + d\left(t - \frac{T}{2}\right)\exp(i(\pi + \theta_{12}))$$

where $d(t)$ is the dipole from a single attosecond burst, $T$ is the laser cycle period and $\theta_{12}$ is the phase difference between these two attosecond bursts. We ignore the amplitude change and consider only the phase difference which can be written as

$$\theta_{12} = \int_{t_1}^{t_{r1}}(\varepsilon(t) - \omega_0)dt - \int_{t_2}^{t_{r2}}(\varepsilon(t) - \omega_0)dt,$$

where $t_i$ and $t_{ri}$ are the time of tunneling and the time when frequency $\omega$ is generated, respectively. $\varepsilon(t)$ is the time-dependent energy difference between the first and second instantaneous eigenstates of the two-level system. So instead of peaking at the odd harmonics, the constructed interference requires that the harmonics peak at the frequency given by

$$\frac{\omega}{\omega_L} = 2n - 1 - \frac{\theta_{12}}{\pi},$$

where $\omega_L$ is the laser frequency, $n$ is integer and $\theta_{12}$ is the energy-dependent phase difference between the two bursts. If $\theta_{12} = 0$ the harmonic photon energies are equal to the odd harmonics of laser photon energy. When the CEP is varied, $\theta_{12}$ is modulated and therefore the harmonic photon energies shift to other frequencies, thus forming slopes as seen in Figure 3a.

**Quantum calculations**

To simulate generation of high-order harmonics in solids, we solve the time-dependent Schrödinger equations in multi-level systems[22] using density matrix approach. For fused silica, we use a system of two levels, where the separation corresponds to the experimental band gap ~9 eV (see inset of Figure 3c). Hamiltonian of the system can be expressed as (in atomic unit (a.u.))

$$\widehat{H}(t) = \begin{bmatrix} 0 & -\mu_0 E(t) \\ -\mu_0 E(t) & \omega_0 \end{bmatrix},$$

where $\omega_0$ is the level separation, $\mu_0$ is the dipole moment between two levels and $E(t)$ is the electric field. We use $\mu_0 = 10$ a.u. and peak field of 2 VÅ$^{-1}$. For crystal quartz, we add one more energy level, which is separated from the second level by ~4 eV corresponding to the spacing of the second conduction band. Also, we include a permanent dipole moment to account for broken inversion symmetry. Therefore, the Hamiltonian becomes:

$$\widehat{H}(t) = \begin{bmatrix} -\mu_p E(t) & -\mu_0 E(t) & 0 \\ -\mu_0 E(t) & \omega_0 & -\mu_1 E(t) \\ 0 & -\mu_1 E(t) & \omega_1 \end{bmatrix},$$

where $\omega_1$ is the separation between first and third level (see Figure 3d), $\mu_p$ is the permanent dipole moment in the valence band, and $\mu_1$ is the dipole moment between second and third level. For quartz we use, $\mu_0 = 5$ a.u., $\mu_1 = 6.6$ a.u. and $\mu_p = 2.5$ a.u. Initially the first level is fully occupied and upper levels are completely empty. The dephasing time is set to 2.8 fs similar to Vampa *et al*[20]. Harmonic spectrum is obtained by Fourier transform of the time-dependent current.

**Acknowledgments:** At Stanford/SLAC this work is supported by the US Department of Energy, Office of Science, Basic Energy Sciences, Chemical Sciences, Geosciences, and Biosciences Division through the Early Career Research Program. The work conducted at UCF is supported by the Air Force Office of Scientific Research under award number FA9550-15-1-0037 and FA9550-16-1-0149, the Army Research Office W911NF-14-1-0383 and W911NF-15-1- 0336, the DARPA PULSE program by a grant from AMRDEC W31P4Q1310017, and the National Science Foundation 1506345. S.G. and Y.S.Y. thank David Reis, Mengxi Wu, Mette Gaarde, Kenneth Schafer, Jerry Hastings and Kelly Gaffney for fruitful discussions.

**Author Contributions:** Y.S.Y. and S.G. conceived the experiments. Y.S.Y., Y.Y., A.C., X.R., M.C., S.G.M., Y.W., F.Z. and S.G. collected and analyzed data. All authors contributed to the interpretation of the results and writing of the manuscript.

**Author Information.** Reprints and permissions information is available at www.nature.com/reprints. The authors declare no competing financial interests. Readers are welcome to comment on the online version of the paper. Correspondence and requests for materials should be addressed to S.G. (shambhu@slac.stanford.edu)

# Figures

**Figure 1 Measured high-harmonic spectra of fused silica and crystal quartz at the peak field ~2 VÅ$^{-1}$**

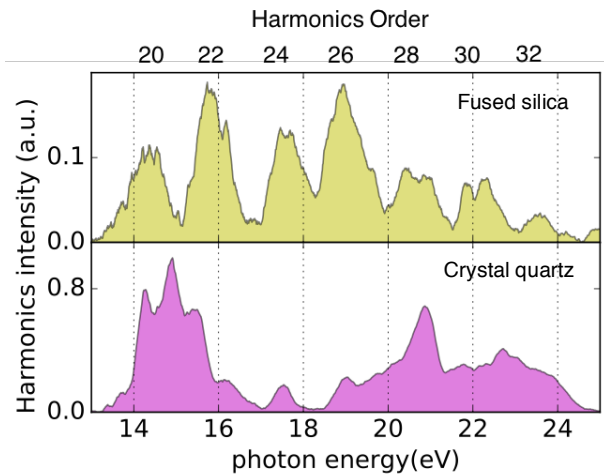

Figure 1 High order harmonics are produced by focusing intense two-cycle laser pulses to ~100 μm thick samples of fused silica and crystal quartz. The central wavelength of the laser is ~1,700 nm (0.73eV). A portion of the spectrum is measured, which shows that the high-energy-end extends to ~25 eV. Both samples withstand repetitive excitation with the maximum peak field of ~2 VÅ$^{-1}$. The spectrum from fused silica consists of discrete harmonic peaks separated by twice the photon energy while the exact location of peaks depends on the CEP setting. The spectrum from crystal quartz shows peaks that separate by one photon energy and merge with each other due to broad spectrum. The efficiency from crystal quartz is higher when it is aligned to the laser field. It also consists of a minimum in the range from 17 eV to 18 eV.

**Figure 2 Dependence of high-harmonic spectrum with the laser peak field**

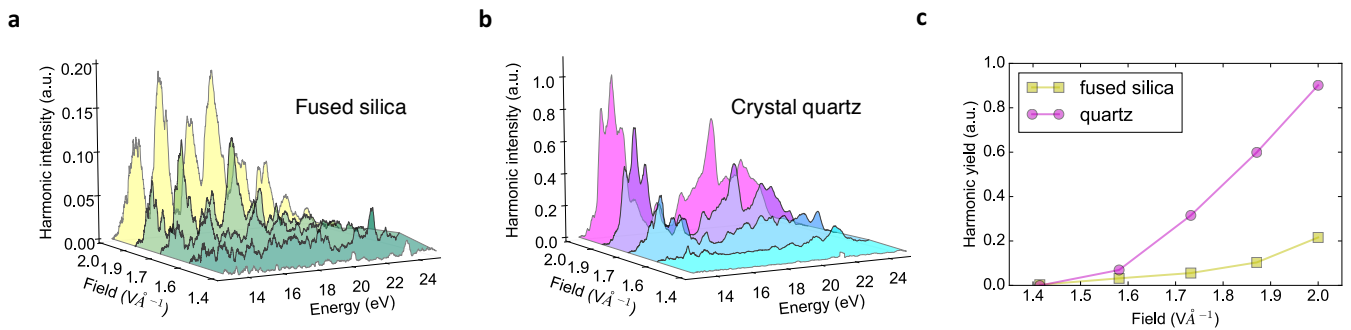

Figure 2 Dependence of high-harmonic spectrum with the peak field of the driving laser for (a) fused silica and (b) crystal quartz. For crystal quartz, the spectral minima seen around 17 eV persists for different peak fields. (c) is the comparison of fused silica and quartz for their total yield (from ~14 to ~25 eV) with different peak fields. At modest fields the total yield is similar but the intensity scaling is significantly different and eventually crystal quartz becomes more efficient.

**Figure 3 Dependence of high-harmonics spectrum with carrier-envelope phase (CEP)**

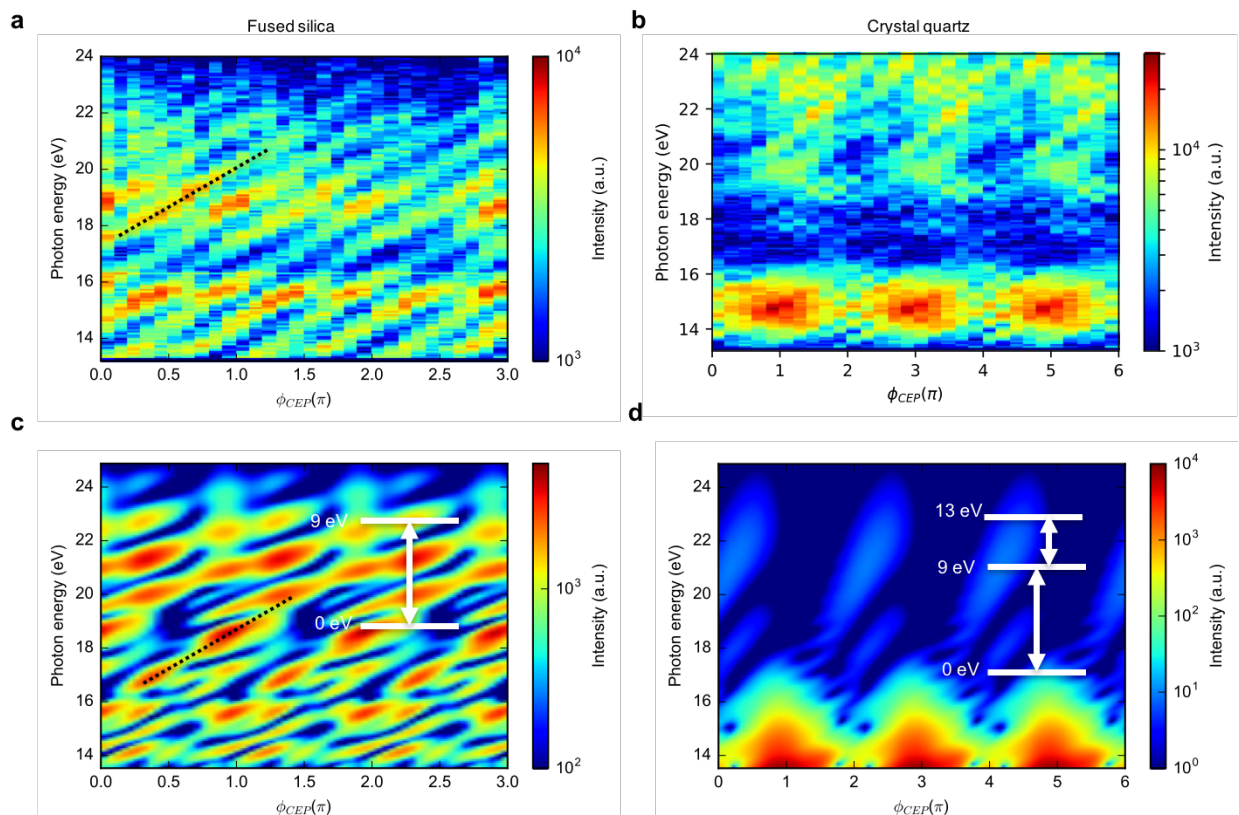

Figure 3 Experimental data for CEP-dependence of high harmonic spectrum from (a) fused silica and (b) crystal quartz at peak laser field of 2 V/Å. Black lines trace the change in photon energy of harmonic peaks with CEP. The amplitude of CEP slope is ~3 eV/π. The spectrum of fused silica repeats every π (horizontal axis) while there is a dominant 2π periodicity for crystal quartz. The spectral minima of crystal quartz persist for all CEP settings. We note that the CEP values in the experiments are relative. (c) and (d) show the calculated high-harmonic spectrum from a quantum mechanical simulation of multi-level models. The insets show the respective energy levels and the couplings used in the simulations. The simulation results reproduce the periodicity and CEP slope of the experimental results.